\documentclass[aps,prl,showpacs,twocolumn,nofootinbib,10pt, superscriptaddress]{revtex4-1}

\usepackage{graphicx}
\usepackage{indentfirst}
\usepackage{subcaption}
\usepackage{parskip}
\usepackage{float}
\usepackage{epstopdf}
\usepackage{amsmath}
\usepackage{amssymb}
\usepackage{mathrsfs}
\usepackage{amsthm}
\usepackage{bm}
\usepackage{url}
\usepackage[T1]{fontenc}
\usepackage{csquotes}
\usepackage[title]{appendix}
\MakeOuterQuote{"}

\newtheoremstyle{note}
  {\topsep/2}               
  {\topsep/2}               
  {}                      
  {\parindent}            
  {\itshape}              
  {.}                     
  {5pt plus 1pt minus 1pt}
  {}

\theoremstyle{note}
\newtheorem{theorem}{Theorem}

\newtheorem{proposition}{Proposition}

\theoremstyle{definition}

\theoremstyle{remark}

\newcommand{\scf}{\mathscr{F}_\mrm{sc}}
\newcommand{\scfp}{\mathscr{F}_\mrm{sc}^+}

\newcommand{\mrm}[1]{\mathrm{#1}}

\newcommand{\gme}{\mathrm{GME}}

\newcommand{\be}{\begin{equation}}
\newcommand{\ee}{\end{equation}}
\newcommand{\ba}{\begin{align}}
\newcommand{\ea}{\end{align}}

\def\<{\langle}  
\def\>{\rangle}  

\newcommand{\cref}[1]{Conjecture~\ref{#1}}
\newcommand{\Cref}[1]{Conjecture~\ref{#1}}

\makeatletter

\newcommand{\Rmnum}[1]{\expandafter\@slowromancap\romannumeral #1@}
\makeatother

\begin{document}
\title{Uniform Construction of Genuine Multipartite Entanglement Measures Based on Geometric Mean and its Applications}
\author{Zong Wang}
\affiliation{School of Mathematical Sciences, MOE-LSC, Shanghai Jiao Tong University, Shanghai 200240,
China}
\affiliation{Shanghai Seres Information Technology Co., Ltd, Shanghai 200040, China}
\affiliation{Shenzhen Institute for Quantum Science and Engineering, Southern University of Science and Technology, Shenzhen 518055, China}
\author{Zhi-Hao Ma}
\email{mazhihao@sjtu.edu.cn}
\affiliation{School of Mathematical Sciences, MOE-LSC, Shanghai Jiao Tong University, Shanghai 200240,
China}
\affiliation{Shanghai Seres Information Technology Co., Ltd, Shanghai 200040, China}
\affiliation{Shenzhen Institute for Quantum Science and Engineering, Southern University of Science and Technology, Shenzhen 518055, China}
\author{Lin Chen}
\email{linchen@buaa.edu.cn}
\affiliation{LMIB(Beihang University), Ministry of Education, and School of Mathematical Sciences, Beihang University, Beijing 100191, China}
\affiliation{International Research Institute for Multidisciplinary Science, Beihang University, Beijing 100191, China}
\author{Cheng-Jie Zhang}
\affiliation{School of Physical Science and Technology, Ningbo University, Ningbo 315211, China}
\email{zhangchengjie@nbu.edu.cn}
\author{Shao-Ming Fei}
\email{smfei@mis.mpg.de}
\affiliation{School of Mathematical Sciences, Capital Normal University, Beijing 100048, China}
\affiliation{Max-Planck-Institute for Mathematics in the Science, Leipzig 04113, Germany}
\date{\today}
\begin{abstract}
Genuine multipartite entanglement (GME) is an important resource in quantum information processing. We systematically study the measures of GME based on the geometric mean of bipartition entanglements. We present a uniform construction of GME measures, which gives rise to the widely used GME measures including GME concurrence, the convex-roof extended negativity of GME, the geometric measure of entanglement of GME. Our GME measures satisfy the desirable conditions such as scalability and smoothness. Moreover, we provide fidelity-based analytical lower bounds for our GME measures. Our bounds are tight and can be estimated experimentally without quantum state tomography. Furthermore, we apply our results to study the dynamics of GME. We identify an initial condition that influences the sudden death of genuine quadripartite entanglement under individual Non-Markovian processes. The GME of Dirac particles with Hawking radiation in the background of a Schwarzschild black hole is also investigated.
\end{abstract}
\maketitle

\section{Introduction}
As a vital resource in quantum information processing, genuine multipartite entanglement (GME) plays a key role in many quantum information tasks such as quantum teleportation \cite{RPMK,CG}, quantum key distribution \cite{RPMK}, quantum cryptography \cite{RPMK,AK} and measurement-based quantum computing \cite{RPMK,RH}. It is also regarded as a useful tool in studying some important physical phenomena such as black hole information paradox \cite{SDM} and quantum phase transition \cite{SMB}.

Detecting and quantifying GME is a challenging task like bipartite entanglement \cite{TFS,YHZ}. There have been some criteria on GME detection \cite{BKS,YXM} some quantifiers of GME such as genuine multipartite concurrence (GMC) \cite{ZHZ}, genuine multipartite negativity \cite{MTO} and three-tangle \cite{YH}. However, both GMC and genuine multipartite negativity are based on the minimal distance between the target state and the bi-separable states, which loses information about the global distribution of entanglement among all parties. The three-tangle is not a proper GME measure because it does not vanish for some bi-separable states. Therefore, it is worth finding better methods to quantify GME.

In \cite{YJ}, the authors proved that the geometric mean of bipartite concurrence (GBC) is a proper measure for GME and has some advantageous properties than other GME measures. In fact, the geometric mean of other bipartite entanglement measures can also be used to quantify GME.
Inspired by this, we propose the geometric mean of convex-roof extended bipartite entanglement measure (GBEM), which is superior to the GME measures based on minimal distance in the sense that it contains also the information about the entanglement distribution among all parties.

Estimation of GME experimentally is difficult for some GME measures due to that the state tomography is inevitable. Recently, the authors in \cite{YYZ} presented lower bounds for some GME measures based on minimal distance. These lower bounds are based on the fidelity between the state and a chosen observable state, avoiding quantum state tomography. Since GBEMs has superiority than the GME measures based on minimal distance, it is of significance to evaluate the GBEMs experimentally without state tomography.

Similar to entanglement sudden death \cite{AFA,MOT,YSJ}, genuine multipartite entanglement sudden death (GMESD) is a kind of dynamical processes in quantum physics, which occurs when the GME is capable of vanishing abruptly. The study on initial conditions that give rise to GMESD has also attracted much attention, which, unlike the entanglement sudden death, is less known up to date. Recently, the authors in \cite{SJ} showed the GMESD in three-qubit systems and found an initial condition which yields the GMESD under certain dynamics.

Quantum information in the background of an asymptotically flat static black hole has been investigated extensively \cite{JQ,ELJ,BQL,QJ,TRR,SXJL}. As one of simple asymptotically flat static black holes, the Schwarzschild black hole was introduced in 1916. Many results have been obtained for quantum entanglement in the background of an asymptotically flat static black hole \cite{JQ,ELJ,BQL,QJ,TRR}. Nevertheless, less is known about the GME in the asymptotically flat static black holes. In \cite{SXJL} the authors investigated the GME of three-qubit states in the background of the Schwarzschild black hole based on 3-tangle \cite{YH}. However, the 3-tangle does not satisfy all the conditions of genuine multipartite entanglement measures \cite{YH}. It is desired to investigate the GME in the background of the Schwarzschild black hole with a proper GME measure.

In this paper, we study the GME based on geometric mean of bipartite entanglement measures in a uniform way and derive their lower bounds in terms of the fidelity between the quantum state and a chosen observable state. In particular, we demonstrate that our lower bounds are tight by detailed example of n-qubit generalized GHZ states. We also show that our results are better than the existing ones \cite{YYZ} for appropriate observable states. Furthermore, we apply our results to study the dynamics of GME, including the genuine 4-partite entanglement sudden death under individual Non-Markovian processes and the GME of Dirac particles with Hawking radiation in the background of a Schwarzschild black hole.

This paper is organized as follows. In Sec. \Rmnum{2}, we present the uniform construction of genuine multipartite entanglement measures, the GBEMs, based on geometric mean. In Sec. \Rmnum{3}, we provide fidelity-based lower bounds for GBEMs and examples to show the efficiency of our results. In Sec. \Rmnum{4}, we apply our results to the dynamics of GME: the genuine 4-partite entanglement under individual Non-Markovian processes, as well as the evolution of GME of Dirac particles with Hawking radiation in the background of a Schwarzschild black hole. We summarize in Sec. \Rmnum{5}.

\section{Uniform construction of GBEMs}
Let us start with the framework of the GBEMs construction. Denote $\scf$ the set of symmetric concave functions on the probability simplex $\vartheta$ with elements $\vartheta_{j}$. Let $\scfp$ be the subset of functions $f$ with the additional property: $f(\vartheta)=0$ iff $\max_{j} \vartheta_j=1$. Note that this condition automatically guarantees that $f(\vartheta)$ is nonnegative. Denote $\gamma|\bar{\gamma}$ a bipartition of a multipartite state. For a pure state $|\varphi\rangle$, we define
\begin{equation}\label{eq:Efgamma}
E^{f}_{\gamma}(|\varphi\rangle):=f(\lambda_{\gamma}(|\varphi\rangle)),
\end{equation}
where $f\in\scfp$, $\lambda_{\gamma}(|\varphi\rangle)$ is the Schmidt vector given by the Schmidt coefficients under the bipartition $\gamma|\bar{\gamma}$.

We declare that for all $f\in\scfp$, $E^{f}_{\gamma}(\rho)$ is an entanglement monotone with convex-roof construction for mixed states $\rho$ \cite{ZH,GV,ZHZ}.
The GBEMs are the geometric means of $E^{f}_{\gamma}(|\psi\rangle)$ over all bipartitions, i.e., for a pure state $|\varphi\rangle$,
\begin{equation}\label{eq:Efgeogme}
G^{f}_\gme(|\varphi\rangle):=(\prod_{\gamma}E^{f}_{\gamma}(|\varphi\rangle))^{\frac{1}{c(\gamma)}},
\end{equation}
where
\begin{equation}\label{Eq:cgamma}
c(\gamma)=
\begin{cases}
\sum_{m=1}^{\frac{n-1}{2}}\tbinom{n}{m}, & if \quad$n is odd$\\[2mm]
\sum_{m=1}^{\frac{n-2}{2}}\tbinom{n}{m}+\frac{1}{2}\tbinom{n}{\frac{n}{2}}, & if \quad$n is even$.\\
\end{cases}
\end{equation}

The GBEMs for mixed states $\rho$ is defined by the convex-roof construction,
\begin{equation}\label{eq:EfmingmeMix}
G^{f}_\gme(\rho):=\min_{\{p_j, |\varphi_j\rangle\}}\sum_{j}p_jG^{f}_\gme(|\varphi_j\rangle),
\end{equation}
where the minimum is taken over all pure-state decompositions of $\rho$.

It is straightforward to verify that $G^{f}_\gme(\rho)$ vanishes when $\rho$ is bi-separable and is positive when $\rho$ is genuine multipartite entangled. $G^{f}_\gme(\rho)$ is also convex by the nature of the convex roof construction \cite{HZZS}, and does not increase under local operation and classical communication (LOCC) by using the concavity of geometric mean function and the Mahler's inequality. Thus GBEMs are well defined GME measures, see the detailed proof in Appendix A.

Furthermore, our GBEMs has the property of smoothness which is absent for those GME measures based on the minimal distance \cite{YJ} such as genuine multipartite concurrence \cite{ZHZ} and genuine multipartite negativity \cite{MTO}. More importantly, the analytical expression of our GBEMs also guarantees that our GBEMs capture all the information about the global distribution of entanglement among all parties.

The GBEMs given by (\ref{eq:EfmingmeMix}) are general in the sense that they include many existing ones as special cases.

(i) If we choose $f(\vartheta)=\sqrt{\frac{d_{min}}{d_{min}-1}\sum_{i\neq j}\vartheta_{i}\vartheta_{j}}$,
$E^{f}_\gme(\rho)$ reduces to the GBC \cite{YJ}, i.e.,
\begin{equation}\label{Eq:GBC}
G_{C}(|\varphi\rangle):=(\prod_{\gamma}C_{\gamma}(|\varphi\rangle))^{\frac{1}{c(\gamma)}},
\end{equation}
where $d_{min}$ represents the dimension of the smaller subsystem under bipartition $\gamma|\bar{\gamma}$ and $C_{\gamma}(|\varphi\rangle)$ is the regularized bipartite concurrence \cite{WKWO}.

(ii) If we take $f(\vartheta)=\sum_{i\neq j}\sqrt{\vartheta_{i}\vartheta_{j}}$, $E^{f}_\gme(\rho)$ reduces to the geometric mean of convex-roof extended negativity $G_{N}(\rho)$, i.e.,
\begin{equation}\label{Eq:GBN}
G_{N}(|\varphi\rangle):=(\prod_{\gamma}N_{\gamma}(|\varphi\rangle))^{\frac{1}{c(\gamma)}},
\end{equation}
where $N_{\gamma}(|\varphi\rangle)$ is the bipartite negativity \cite{GR,SDS} under bipartition $\gamma|\bar{\gamma}$.

(iii) If we set $f(\vartheta)=m_{\gamma}(\prod_{i=1}^{m_{\gamma}}\vartheta_{i})^{\frac{1}{m_{\gamma}}}$, $E^{f}_\gme(\rho)$ gives rise to the geometric mean of bipartite G-concurrence $G_{GC}(\rho)$, i.e.,
\begin{equation}\label{Eq:GBGC}
G_{GC}(|\varphi\rangle):=(\prod_{\gamma}\textit{GC}_{\gamma}(|\varphi\rangle))^{\frac{1}{c(\gamma)}},
\end{equation}
where $\textit{GC}_{\gamma}(|\varphi\rangle)$ is the G-concurrence \cite{GGU,HKH} and $m_{\gamma}$ is the number of non vanishing Schmidt coefficients under the bipartition $\gamma|\bar{\gamma}$.

(iv) If we let $f(p)=1-\max_{i}{\vartheta_{i}}$, $E^{f}_\gme(\rho)$ becomes the geometric mean of geometric measure of entanglement, i.e.,
\begin{equation}\label{Eq:GBGM}
G_{GM}(|\varphi\rangle):=(\prod_{\gamma}\textit{GM}_{\gamma}(|\varphi\rangle))^{\frac{1}{c(\gamma)}},
\end{equation}
where $\textit{GM}_{\gamma}(|\varphi\rangle)$ is the geometric measure of bipartite entanglement \cite{HNL,TCPM} under the bipartition $\gamma|\bar{\gamma}$.

\section{Fidelity-based Lower Bounds of GBEMs}

To evaluate the GBEMs, we first give an improved lower bound of the geometric mean of bipartite concurrence given by (\ref{Eq:GBC}), based on the fidelity between the target state and an arbitrarily chosen pure state $|\psi\rangle$.

Consider the Schmidt decomposition of an n-partite pure state $|\psi\rangle$ under bipartition $\gamma|\bar{\gamma}$,
\begin{equation}\label{Eq:SD}
|\psi\rangle=\sum_{i=0}^{m_{\gamma}-1}\sqrt{\lambda_{i}^{\gamma}}|ii\rangle,
\end{equation}
where $\{\sqrt{\lambda_{i}^{\gamma}}\}_{i=0}^{m_{\gamma}-1}$ are the Schmidt coefficients in decreasing order, $m_{\gamma}$ is the corresponding Schmidt rank. For simplicity we define
\begin{equation}\label{Eq:lamdamd1}
\lambda_{0}^{(1)}=\max_{\gamma}{\lambda_{0}^{\gamma}},~
m^{(1)}=\max_{\gamma}{m_{\gamma}},~
d_{min}^{(1)}=\max_{\gamma}{d_{min}^{\gamma}},
\end{equation}
where $d_{min}^{\gamma}$ is the dimension of the smaller subsystem under bipartition $\gamma|\bar{\gamma}$. Similarly we denote $\lambda_{0}^{(2)}$, $m^{(2)}$ and $d_{min}^{(2)}$ the second largest of $\lambda_{0}^{\gamma}$, $m^{\gamma}$ and $d_{min}^{\gamma}$ over all bi-partitions, respectively. We define
\begin{equation}\label{Eq:Lamdaj}
\Lambda_{\psi}^{(j)}=\max\{1,\frac{\langle\psi|\rho|\psi\rangle}{\lambda_{0}^{(j)}}\}
\end{equation}
and
\begin{equation}\label{Eq:A1A2}
A_{\psi}^{(j)}=\sqrt{\frac{d_{min}^{(j)}}{(d_{min}^{(j)}-1)m^{(j)}(m^{(j)}-1)}}(\Lambda_{\psi}^{(j)}-1)
\end{equation}
with $j\in\{1,2\}$.

\begin{theorem}
For an arbitrary n-partite state $\rho$, the geometric mean of bipartite concurrence $\textit{G}_{C}(\rho)$ satisfies
\begin{equation}\label{eq:lbGBC}
\textit{G}_{C}(\rho)\geq \max_{|\psi\rangle}[A_{\psi}^{(1)}(A_{\psi}^{(2)})^{c(\gamma)-1}]^{\frac{1}{c(\gamma)}},
\end{equation}
where $c(\gamma)$ and $A^{(j)}$ $(j=1,2)$ are defined in Eq.\eqref{Eq:cgamma} and Eq.\eqref{Eq:A1A2}, respectively.
\end{theorem}

The proof is in Appendix B. Although our lower bound is given by taking over all pure observable states $|\psi\rangle$, it is worth mentioning that $[A_{\psi}^{(1)}(A_{\psi}^{(2)})^{c(\gamma)-1}]^{\frac{1}{c(\gamma)}}$ is also a lower bound for any selected $|\psi\rangle$. Moreover, $\lambda_{0}^{(j)}$, $m^{(j)}$ and $d_{min}^{(j)}$ are given for any fixed $|\psi\rangle$. For example, if we choose the n-qubit Greenberger-Horne-Zeilinger (GHZ) state $|GHZ_{n}\rangle=\frac{1}{\sqrt{2}}\sum_{i=0}^{1}|ii...i\rangle$ as the observable state, then we have $\lambda_{0}^{(1)}=\lambda_{0}^{(2)}=\frac{1}{2}$, $m^{(1)}=m^{(2)}=2$ and $d_{min}^{(1)}=d_{min}^{(2)}=2$. The lower bound is obtained by experimentally measuring the fidelity between $\rho$ and the chosen observable state $|\psi\rangle$.
We give two examples below to show that our bound is tight for some states and better than the results in \cite{YYZ}.

\textit{Example 1}
To illustrate the tightness of our bound, we consider the n-qudit generalized GHZ state, $|\psi_{GHZ_{n}}\rangle=\sum_{i=0}^{d-1}c_{i}|ii...i\rangle$, where $c_{i}$s are all real. We choose the n-qudit GHZ state $|GHZ_{nd}\rangle=\frac{1}{\sqrt{d}}\sum_{i=0}^{d-1}|ii...i\rangle$ as the observable state. By direct calculation we have
\begin{equation*}
\textit{G}_{C}(|\psi_{GHZ_{n}}\rangle)=
\prod_{m=1}^{\frac{n-1}{2}}\left(\frac{2d^{m}\,\Omega}{d^{m}-1}\right)^\eta
\end{equation*}
if $n$ is odd, and
\begin{equation*}
\textit{G}_{C}(|\psi_{GHZ_{n}}\rangle)=
\left(\frac{d^{\frac{n}{2}}\,\Omega}{d^{\frac{n}{2}}-1}\right)^{\frac{1}{2}}
\prod_{m=1}^{\frac{n-2}{2}}\left(\frac{2d^{m}\,\Omega}{d^{m}-1}\right)^\eta
\end{equation*}
if $n$ is even, where $\Omega=\sum_{i<j}c_{i}^{2}c_{j}^{2}$ and $\eta=\tbinom{n}{m}/2$, which exactly equal to our lower bounds. When $d=2$ and $c_{0}=c_{1}=\frac{1}{\sqrt{2}}$, our lower bound recovers the one given in the Appendix B of \cite{YJ}.

\textit{Example 2}
For the 3-qubit state $\rho_{2}=p|W_{3}\rangle\langle W_{3}|+\frac{1-p}{8}I$ with $|W_{3}\rangle=\frac{1}{\sqrt{3}}|100\rangle+|010\rangle+|001\rangle$, we choose $|\phi\rangle=\frac{1}{2}(|100\rangle+|010\rangle)+\frac{1}{\sqrt{2}}|001\rangle$ as the observable state whose $\lambda^{(1)}=\frac{3}{4}$ and $\lambda^{(2)}=\frac{1}{2}$.
One gets that $\rho_{2}$ is genuine tripartite entangled when $p\in(0.7384,1]$ by using the lower bound in \cite{YYZ}. Our result include the results in \cite{YYZ} and can also ensure that $\rho_{2}$ is genuine tripartite entangled when $p\in(0.4431,0.7384)$. It demonstrates that our result detects more genuine entangled states than the results in \cite{YYZ}.
Specially, our result guarantees that $\rho_{2}$ is genuine tripartite entangled when $p\in(0.6190,1]$ if we choose $|W_{3}\rangle$ as the observable state, which is equal to the result in \cite{YYZ}.

Similar to Theorem 1, we have the following lower bounds for the geometric mean for convex-roof extended negativity, G-concurrence and geometric measure of entanglement for n-partite pure states given by \eqref{Eq:GBN}-\eqref{Eq:GBGM}, respectively, see proofs in Appendix C.

\begin{proposition}
For an arbitrary n-partite state $\rho$, the lower bound of the geometric mean of convex-roof extended negativity $\textit{G}_{N}(\rho)$ is given by
\begin{equation}\label{eq:lbGBN}
\textit{G}_{N}(\rho)\geq \max_{|\psi\rangle}[B_{\psi}^{(1)}(B_{\psi}^{(2)})^{c(\gamma)-1}]^{\frac{1}{c(\gamma)}},
\end{equation}
where
$B_{\psi}^{(j)}=\Lambda_{\psi}^{(j)}-1$, $j\in\{1,2\}$.
\end{proposition}

\begin{proposition}
For an arbitrary n-partite state $\rho$, the lower bound of the geometric mean of G-concurrence $\textit{G}_{GC}(\rho)$ satisfies
\begin{equation}\label{eq:lbGBGC}
\textit{G}_{GC}(\rho)\geq \max_{|\psi\rangle}[C_{\psi}^{(1)}(C_{\psi}^{(2)})^{c(\gamma)-1}]^{\frac{1}{c(\gamma)}},
\end{equation}
where
$C_{\psi}^{(j)}=[1-m^{(j)}+(\Lambda_{\psi}^{(j)}-1)]$, $j\in\{1,2\}$.
\end{proposition}

\begin{proposition}
For an arbitrary n-partite state $\rho$, the geometric mean of geometric measure of bipartite entanglement $\textit{G}_{GC}(\rho)$ satisfies
\begin{equation}\label{eq:lbGGME}
\textit{G}_{GM}(\rho)\geq \max_{|\psi\rangle}[D_{\psi}^{(1)}(D_{\psi}^{(2)})^{c(\gamma)-1}]^{\frac{1}{c(\gamma)}},
\end{equation}
where
$D_{\psi}^{(j)}=[1-\frac{[\sqrt{\Lambda_{\psi}^{(j)}}+\sqrt{(m^{(j)}-1)(m^{(j)}-\Lambda_{\psi}^{(j)})}]^{2}}{(m^{(j)})^{2}}]^{\frac{1}{c(\gamma)}},$
$j\in\{1,2\}$.
\end{proposition}

For any n-qubit GHZ state $|GHZ_{n}\rangle$, the bounds given in proposition 1-3 are equal to the true values by choosing the observable state to be $|GHZ_{n}\rangle$ itself.
As a special case, our bounds are equal to the results in \cite{YYZ} when the observable state satisfies $\lambda^{(1)}=\lambda^{(2)}$, for instance, the $|GHZ_{n}\rangle$ state and the n-qubit W state $|W_{n}\rangle=\frac{1}{\sqrt{n}}\sum_{j=1}^{n}|0_{1}...0_{j-1}1_{j}0_{j+1}...0_{n}\rangle$. For other observable states satisfying $\lambda^{(1)}>\lambda^{(2)}$, our lower bounds are larger than the bounds in \cite{YYZ}.


\section{Dynamics Of Genuine Multipartite Entanglement}

\noindent{\it Genuine 4-partite entanglement sudden death via individual non-Markovian process}~~
Let us start with the initial four-qubit system state,
\begin{equation*}
|\psi_{GHZ_{4}}\rangle=\cos\alpha|0000\rangle+\sin\alpha|1111\rangle,
\end{equation*}
and the environmental state $|0000\rangle$, where $\alpha\in[0,\frac{\pi}{2}]$.
Consider the individual non-Markovian process that the interaction between the individual system qubits and the non-Markovian hot storage (environment) is described by the following Hamiltonian,
\begin{equation}\label{MK}
H_{Aa}=\omega_{0}\sigma_{+}^{A}\sigma_{-}^{A}+\sum_{i}\omega_{i}k_{i}^{+}k_{i}+\sum_{i}(g_{i}\sigma_{+}^{A}k_{i}+g_{i}^{+}\sigma_{-}^{A}k_{i}^{+}),
\end{equation}
where $A$ and $a$ denote the first system qubit and the environmental (hot storage) qubit, respectively, $\sigma_{\pm}^{A}$ is the lift operator of system $A$ and $\omega_{0}$ is the conversion frequency, $\omega_{i}$ and $k_{i}^{+}(k_{i})$ are the frequency and creation (annihilation) operators of the $i$-th mode of the $k$-th hot storage respectively, $g_{i}$ represents the coefficients of the coupling between system $A$ and the $i$-th mode of the hot storage.

\begin{figure}[htp]
  \includegraphics[width=6cm]{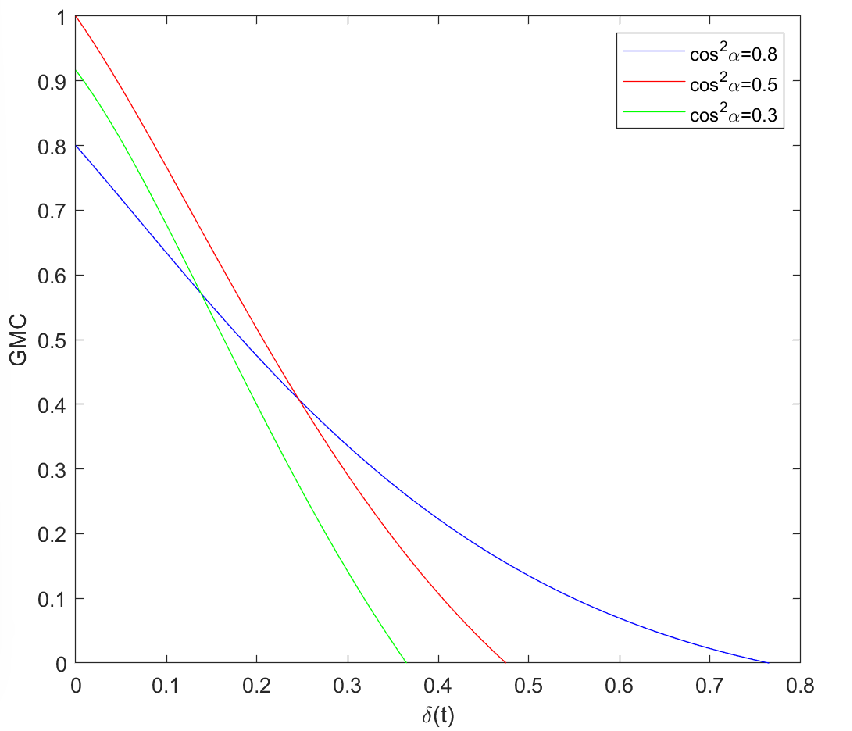}
\caption{The evolution of the GMC of the system state.}
\label{Fig.3}
\end{figure}

When the Markovian decay rate is larger than half of the spectrum width of the coupling in zero temperature, the individual Non-Markovian processes can be described by (\ref{MK}). The individual Non-Markovian process is described by \cite{BRG}
\begin{eqnarray*}\label{MKMATH}
&&|0_{A}0_{a}\rangle\longrightarrow|0_{A}0_{a}\rangle,\\
&&|1_{A}0_{a}\rangle\longrightarrow\sqrt{p}|1_{A}0_{a}\rangle+\sqrt{1-p}|0_{A}1_{a}\rangle,
\end{eqnarray*}
which is similar to the individual amplitude damping channel \cite{SJ}, where $p=e^{-\delta(t)}$ is a time $t$ dependent coefficient.
The density matrix of system state via individual Non-Markovian process is an X-matrix. We can calculate its GMC \cite{SMCJ} and get that the genuine 4-qubit entanglement sudden death occurs when $|\cot\alpha|>7(1-p)^{2}$. Moreover, we find that the sudden death happens earlier for larger $\alpha$, see Fig. \ref{Fig.3}. It implies that the portion of $|1111\rangle$ in the initial state makes the occurrence of genuine 4-qubit sudden death earlier, which is a generalization of the result in \cite{SJ}.

For comparison, we also calculate our lower bound of GBC for the system state. It is shown that for the same $\alpha$ given in Fig. \ref{Fig.3}, lower bound of GBC changes with $\delta(t)$ in accordance with the behavior of GMC, by choosing $|GHZ_{4}\rangle$ as the observable state.

\begin{figure}[htp]
\centering
\begin{minipage}[t]{0.5\linewidth}
  \centering
  \includegraphics[width=1\textwidth]{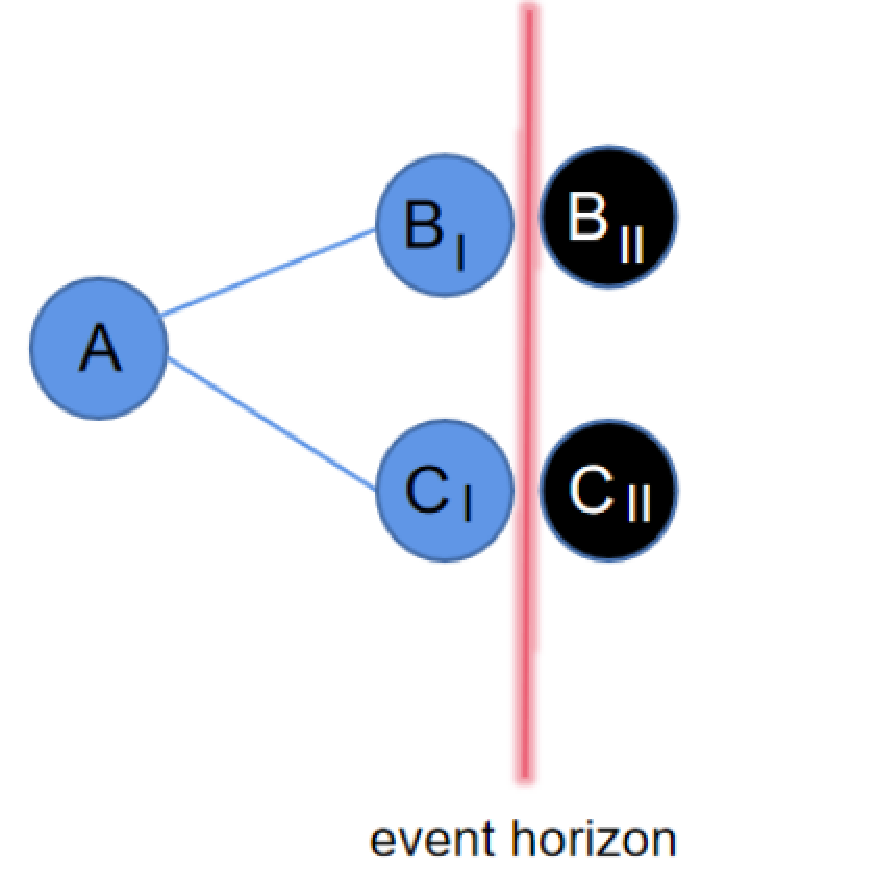}
  \centerline{(a)}
\end{minipage}%
\begin{minipage}[t]{0.5\linewidth}
  \centering
  \includegraphics[width=1\textwidth]{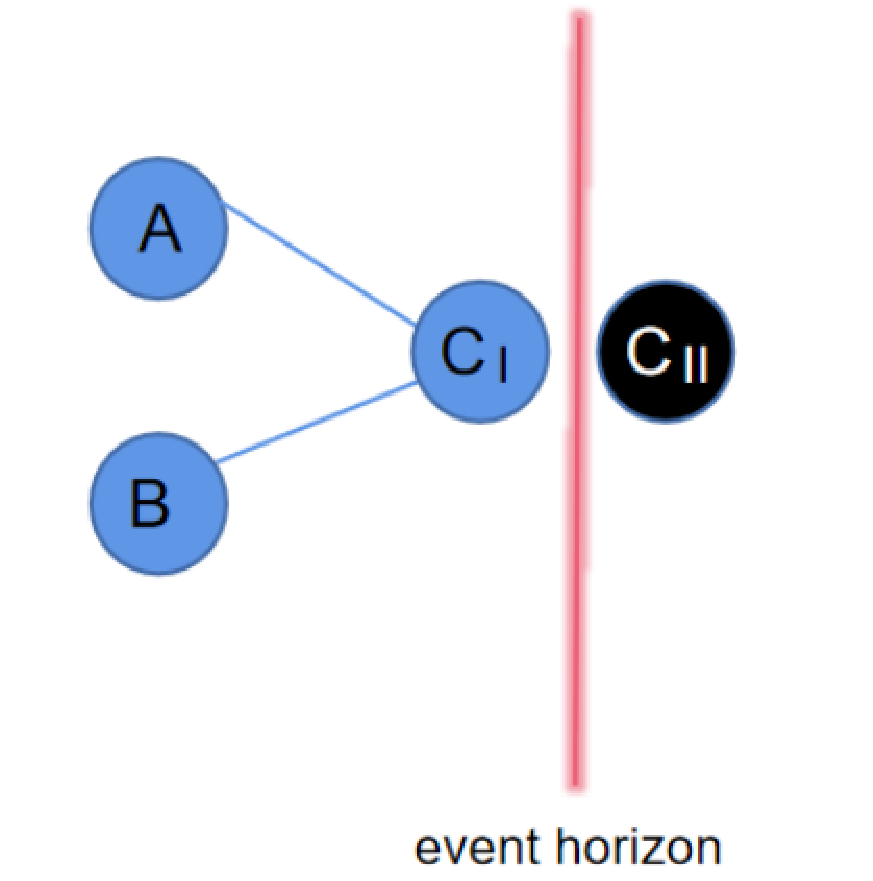}
  \centerline{(b)}
\end{minipage}
\caption{Two cases in the background of a Schwarzschild black hole.}
\label{Fig.4}
\end{figure}

\noindent{\it The effect of Hawking Radiation on GME of Dirac particles in the background of a Schwarzschild black hole}~~
Consider that Alice, Bob and Charlie share a three-qubit generalized GHZ state $|\psi_{GHZ_{3}}\rangle=cos\theta|000\rangle+sin\theta|111\rangle$ in the inertial Minkowski space-time \cite{SXJL} with $\theta\in[0,\frac{\pi}{2}]$. Each has a particle detector to detect their own mode. One or two of them fall freely into the black hole along the geodesic line \cite{CWD}, with spinning and accelerating around the black hole near the event horizon. The rest of them are (is) still in the asymptotically flat region, as shown in Fig. \ref{Fig.4}.

Fig. \ref{Fig.4}(a) shows that Bob and Charlie fall freely into the black hole while Alice is still in the asymptotically flat region, which means that Bob and Charlie are in black hole mode and Alice in Minkowski mode. According to the results in \cite{TRR} the shared state becomes
\begin{equation*}
\begin{split}
|\psi'_{G}\rangle=&cos\theta(e^{-\frac{\omega}{T}}+1)^{-1}|0_{A}0_{B_{1}}
0_{B_{2}}0_{C_{1}}0_{C_{2}}\rangle\\
&+cos\theta(e^{\frac{\omega}{T}}+1)^{-1}|0_{A}1_{B_{1}}1_{B_{2}}1_{C_{1}}1_{C_{2}}\rangle\\
&+sin\theta|1_{A}1_{B_{1}}0_{B_{2}}1_{C_{1}}0_{C_{2}}\rangle\\
&+cos\theta(e^{-\frac{\omega}{T}}+e^{-\frac{\omega}{T}}+2)^{-\frac{1}{2}}
|0_{A}0_{B_{1}}0_{B_{2}}1_{C_{1}}1_{C_{2}}\rangle\\
&+cos\theta(e^{-\frac{\omega}{T}}+e^{-\frac{\omega}{T}}+2)^{-\frac{1}{2}}
|0_{A}1_{B_{1}}1_{B_{2}}0_{C_{1}}0_{C_{2}}\rangle,
\end{split}
\end{equation*}
where $T$ and $\omega$ are the Hawking temperature and Dirac field frequency, respectively.
$X_{1}$ and $X_{2}$ represent the subsystems beyond and within the event horizon, respectively, for $X\in\{A,B,C\}$. Due to the causal independence between the inside and outside of the event horizon, one can only obtain the entanglement among $A$, $B_{1}$ and $C_{1}$.

\begin{figure}
  \centering
  \includegraphics[width=6cm]{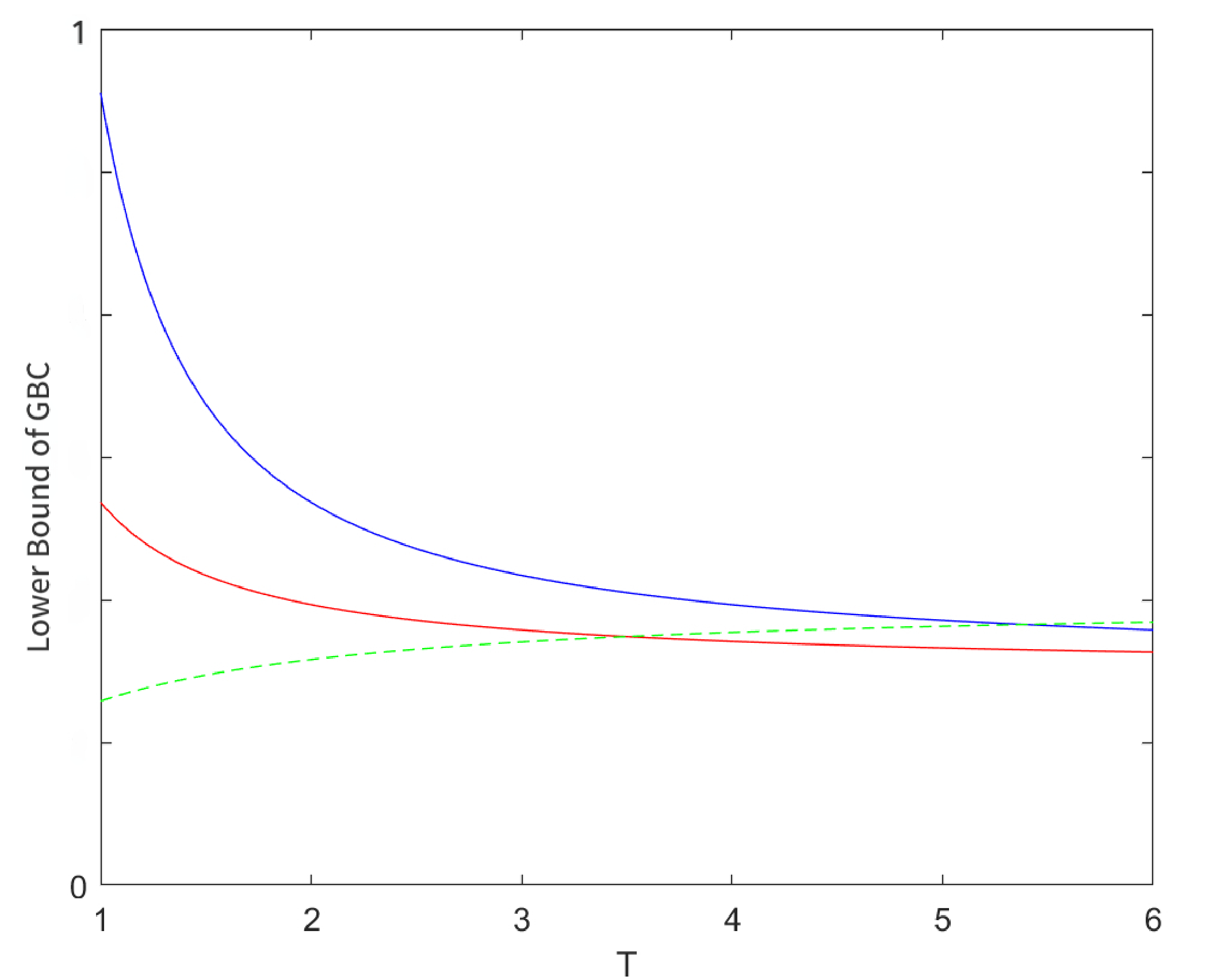}\\
  \caption{The Lower Bound of GBC versus Hawking temperature. The blue solid line represents the lower bound of the obtainable GMC as a function of the Hawking temperature for the case of Fig. \ref{Fig.4}(a). The red solid and green dashed lines represent the obtainable and unobtainable GMC for the case of Fig. \ref{Fig.4}(b), respectively.}\label{Fig.5}
\end{figure}

The system state after the process in Fig. \ref{Fig.4} is not an X-matrix. The corresponding GMC can not be calculated analytically. Instead we calculate the lower bound of GBC given by the Theorem 1. Set $\omega=1$ and $\theta=\frac{\pi}{4}$. The relation between the lower bound of the obtainable GBC and the Hawking temperature is shown in (blue solid line) Fig. \ref{Fig.5}.

With respect to Fig. \ref{Fig.4}(b), where Alice and Bob are still in the asymptotically flat region, while Charlie falls freely into the black hole, the shared state becomes \cite{TRR}
\begin{equation*}
\begin{split}
|\psi''_{G}\rangle=&cos\theta(e^{-\frac{\omega}{T}}+1)^{-\frac{1}{2}}|0_{A}0_{B}0_{C_{1}}0_{C_{2}}\rangle\\
&+cos\theta(e^{\frac{\omega}{T}}+1)^{-\frac{1}{2}}|0_{A}0_{B}1_{C_{1}}1_{C_{2}}\rangle\\
&+sin\theta|1_{A}1_{B}1_{C_{1}}0_{C_{2}}\rangle.
\end{split}.
\end{equation*}
The causal independence between the inside and outside of the event horizon guarantees that only the GME among subsystems $A, B, C_{1}$ is obtainable. In contrast, the GME among $A, B, C_{2}$ cannot be obtained. We estimate these two kinds of GBC as a function of the Hawking temperature with the same $\omega$ and $\theta$ as above, denoted by the red solid line and green dashed line in Fig. \ref{Fig.5}, respectively. It is seen that the GBC among $A, B, C_{1}$ reduces to a constant with the decreased frequency of descent, which is opposite to the GBC among $A, B, C_{2}$. It implies an interesting phenomenon that a fraction of the decreased obtainable GBC is transformed into the fraction of the unobtainable GBC.

\section{Summary}
We have systematically studied the genuine multipartite entanglement measures based on geometric mean of bipartite entanglement measures, and presented unform constructions of the
genuine multipartite entanglement measures. The fidelity-based lower bounds of some typical GBEMs from such constructions are derived analytically. We have demonstrated that our bounds are tight and more powerful than the existing ones \cite{YYZ} by two detailed examples. Furthermore, we have investigated the dynamics of GME. An initial condition that influences the sudden death of genuine 4-partite entanglement has been found under individual Non-Markovian processes. Moreover, we have examined the GME for Dirac particles with Hawking radiation in the background of a Schwarzschild black hole. Our results may be also applied to investigate other dynamics of GME and highlight further researches on GBEMs in different physical systems.

\section{Acknowledgments}
This work is supported by Fundamental Research Funds for the Central Universities; NSFC (Grant Nos. 12075159, 12171044 and 12371132); Shenzhen Institute for Quantum Science and Engineering, Southern University of Science and Technology (Grant Nos. SIQSE202005); the specific research fund of the Innovation Platform for Academicians of Hainan Province; NNSF of China (Grant Nos. 11734015 and 11871089).

\section{APPENDIX A: Proof: the GBEMs are GME measures}
A well defined GME measure $E(\rho)$ should satisfies

(C1)\begin{equation*}
E(\rho)=0
\end{equation*}
for any biseparable state $\rho$.

(C2)\begin{equation*}
E(\rho)>0
\end{equation*}
for any genuine multipartite entangled state $\rho$.

(C3)\begin{equation*}
E(\sum_{j}p_{j}|\psi_{j}\rangle)\leq\sum_{j}p_{j}E(|\psi_{j}\rangle).
\end{equation*}

(C4)\begin{equation*}
E(\Lambda(\rho))\leq E(\rho)
\end{equation*}
for LOCC $\Lambda$

$G^{f}_\gme(\rho)$ is obviously convex by the nature of the convex roof construction. In addition, one can easily get that $G^{f}_\gme$ vanish when $\rho$ is biseparable and positive when $\rho$ is genuine multipartite entangled for all $f\in\scfp$. To show $G^{f}_\gme(\rho)$ satisfies condition (C4), we prove it does not increase under selective LOCC \cite{HZZS} first. We assume that the initial state $\rho$ is pure and there is a given LOCC transforms $\rho$ to $\rho_j$ with probability $p_j$, by using the concavity of geometric mean function and Mahler's inequality, we can get
\begin{equation*}
\begin{split}
G^{f}_\gme(\rho):=(\prod_{\gamma}E^{f}_{\gamma}(\rho))^{\frac{1}{c(\gamma)}}\geq (\prod_{\gamma}\sum_{j}p_{j}E^{f}_{\gamma}(\rho_{j}))^{\frac{1}{c(\gamma)}}\\
\geq\sum_{j}p_{j}(\prod_{\gamma}E^{f}_{\gamma}(\rho_{j}))^{\frac{1}{c(\gamma)}}=\sum_{j}p_{j}G^{f}_\gme(\rho_{j}).
\end{split}
\end{equation*}
Therefore, $G^{f}_\gme(\rho)$ does not increase under selective LOCC. Combined with condition (C3), one can get $G^{f}_\gme(\rho)$ satisfies condition (C4) \cite{HZZS}, which completes the proof.

\section{APPENDIX B: Proof Of Theorem 1}
\begin{proof}
We suppose $\rho=\sum_{k}p_{k}|\varphi_{k}\rangle\langle\varphi_{k}|$ is the optimal decomposition of an arbitrary mixed state $\rho$, i.e., $\textit{G}_{C}(\rho)=\sum_{k}p_{k}\textit{G}_{C}(|\varphi_{k}\rangle)$. For any bipartition $\gamma|\bar{\gamma}$, the regulized concurrence of a pure state is defined by
\begin{equation*}
\begin{split}
C_{\gamma}(|\varphi_{k}\rangle)&=\sqrt{\frac{d_{min}}{d_{min}-1}[1-tr((\rho_{k})_{\gamma}^{2})]}\\
&\geq
\sqrt{\frac{d_{min}^{(j)}}{(d_{min}^{(j)}-1)m^{(j)}(m^{(j)}-1)}}(\frac{\langle\psi|\rho|\psi\rangle}{\lambda_{0}^{\gamma}}-1),
\end{split}
\end{equation*}
where $\rho_{k}=|\varphi_{k}\rangle\langle\varphi_{k}|$ and the second inequality is proved in \cite{YYZ}. Thus we get for any $|\psi\rangle$,
\begin{equation*}
\begin{split}
\textit{G}_{C}(|\varphi_{k}\rangle)&=(C_{\gamma}(|\varphi_{k}\rangle))^{\frac{1}{c(\gamma)}}\\
&\geq [\prod_{j=1}^{c(\gamma)}\sqrt{\frac{d_{min}^{(j)}}{(d_{min}^{(j)}-1)m^{(j)}(m^{(j)}-1)}}(\Lambda_{\psi}^{(j)}-1)]^{\frac{1}{c(\gamma)}}\\
&\geq
[A_{\psi}^{(1)}(A_{\psi}^{(2)})^{c(\gamma)-1}]^{\frac{1}{c(\gamma)}}.
\end{split}
\end{equation*}
Therefore,
\begin{equation*}
\begin{aligned}
\textit{G}_{C}(\rho)&=\sum_{k}p_{k}\textit{G}_{C}(|\varphi_{k}\rangle)\\
&\geq [A_{\psi}^{(1)}(A_{\psi}^{(2)})^{c(\gamma)-1}]^{\frac{1}{c(\gamma)}}
\end{aligned}
\end{equation*}
holds for any $|\psi\rangle$, which completes the proof of Theorem 1.
\end{proof}

\section{APPENDIX C: Proofs of Proposition 1-3}
\subsection{Proof of Proposition 1}
\begin{proof}
For the same optimal decomposition of $\rho$ in the proof of Theorem 1, the bipartite negativity of a pure state is defined by
\begin{equation*}
\begin{split}
N_{\gamma}(|\varphi_{k}\rangle)&=\||\varphi_{k}\rangle\langle\varphi_{k}|^{T_{B}}\|_{tr}-1\\
&\geq
\frac{\langle\psi|\rho|\psi\rangle}{\lambda_{0}^{\gamma}}-1,
\end{split}
\end{equation*}
where $X^{T_{B}}$ and $\|X\|_{tr}$ represent the partial transpose and trace norm of $X$, respectively. The second inequality is proposed in \cite{YYZ}. Thus can get for any $|\psi\rangle$,
\begin{equation*}
\begin{split}
\textit{G}_{N}(|\varphi_{k}\rangle)&=(N_{\gamma}(|\varphi_{k}\rangle))^{\frac{1}{c(\gamma)}}\\
&\geq [\prod_{j=1}^{c(\gamma)}(\Lambda^{(j)}-1)]^{\frac{1}{c(\gamma)}}\\
&\geq
[B_{\psi}^{(1)}(B_{\psi}^{(2)})^{c(\gamma)-1}]^{\frac{1}{c(\gamma)}}.
\end{split}
\end{equation*}
According to this, one can get for any $|\psi\rangle$,
\begin{equation*}
\begin{aligned}
\textit{G}_{N}(\rho)&=\sum_{k}p_{k}\textit{G}_{N}(|\varphi_{k}\rangle)\\
&\geq [B_{\psi}^{(1)}(B_{\psi}^{(2)})^{c(\gamma)-1}]^{\frac{1}{c(\gamma)}}.
\end{aligned}
\end{equation*}
The proof of Proposition 1 is completed.
\end{proof}

\subsection{Proof of Proposition 2}
\begin{proof}
For the same optimal decomposition of $\rho$ in the proof of Theorem 1, the bipartite G-concurrence of a pure state is defined by
\begin{equation*}
\begin{aligned}
\textit{GC}_{\gamma}(|\varphi_{k}\rangle)&=d(\gamma)(det[(\rho_{k})_{\gamma}])^{\frac{1}{d(\gamma)}}\\
&\geq
1-m_{\gamma}+\frac{\langle\psi|\rho|\psi\rangle}{\lambda_{0}^{\gamma}},
\end{aligned}
\end{equation*}
where $\rho_{k}=|\varphi_{k}\rangle\langle\varphi_{k}|$, and $d(\gamma)$ is the dimension of the smaller subsystem under bipartition $\gamma|\bar{\gamma}$. The second inequality is proposed in \cite{YYZ}. Thus we get for any $|\psi\rangle$,
\begin{equation*}
\begin{split}
\textit{G}_{GC}(|\varphi_{k}\rangle)&=(\textit{GC}_{\gamma}(|\varphi_{k}\rangle))^{\frac{1}{c(\gamma)}}\\
&\geq [\prod_{j=1}^{c(\gamma)}(1-m_{\gamma}+\frac{\langle\psi|\rho|\psi\rangle}{\lambda_{0}^{\gamma}})]^{\frac{1}{c(\gamma)}}\\
&\geq
[C_{\psi}^{(1)}(C_{\psi}^{(2)})^{c(\gamma)-1}]^{\frac{1}{c(\gamma)}}.
\end{split}
\end{equation*}
Based on this, one can get for any $|\psi\rangle$,
\begin{equation*}
\begin{aligned}
\textit{G}_{GC}(\rho)&=\sum_{k}p_{k}\textit{G}_{GC}(|\varphi_{k}\rangle)\\
&\geq [C_{\psi}^{(1)}(C_{\psi}^{(2)})^{c(\gamma)-1}]^{\frac{1}{c(\gamma)}}.
\end{aligned}
\end{equation*}
Thus we complete the proof of Proposition 2.
\end{proof}

\subsection{Proof of Proposition 3}
\begin{proof}
For a pure state $|\varphi_{k}\rangle$, the geometric measure of bipartite entanglement is defined by
\begin{equation*}
\textit{GM}_{\gamma}(|\varphi_{k}\rangle)=1-\max_{i}\{\nu_{i}\},
\end{equation*}
where $\sqrt{\nu_{i}}$ are the Schmidt coefficients of $\rho_{k}$ in decreasing order under bipartition $\gamma|\bar{\gamma}$, which is similar to Eq. (\ref{Eq:SD}). Then we have for any $|\psi\rangle$,
\begin{equation*}
\begin{split}
\textit{G}_{GM}(&|\varphi_{k}\rangle)
=(\textit{GM}_{\gamma}(|\varphi_{k}\rangle))^{\frac{1}{c(\gamma)}}\\
&\geq (\prod_{j=1}^{c(\gamma)}(1-\frac{[\sqrt{\Lambda^{(j)}}+\sqrt{(m^{(j)}-1)(m^{(j)}
-\Lambda^{(j)})}]^{2}}{(m^{(j)})^{2}}))^{\frac{1}{c(\gamma)}}\\
&\geq
[D_{\psi}^{(1)}(D_{\psi}^{(2)})^{c(\gamma)-1}]^{\frac{1}{c(\gamma)}}.
\end{split}
\end{equation*}
The second inequality is proposed in \cite{YYZ}. Therefore, one get for any $|\psi\rangle$,
\begin{equation*}
\begin{aligned}
\textit{G}_{GM}(\rho)&=\sum_{k}p_{k}\textit{G}_{GM}(|\varphi_{k}\rangle)\\
&\geq [D_{\psi}^{(1)}(D_{\psi}^{(2)})^{c(\gamma)-1}]^{\frac{1}{c(\gamma)}}.
\end{aligned}
\end{equation*}
The proof of Proposition 3 is completed.
\end{proof}

\end{document}